\documentclass[aps,nofootinbib]{revtex4}
\usepackage{color,epsfig,graphics,amssymb,amsmath,subeqnarray,scalerel}

\def\d{{\rm d}}\def\p{\partial}\def\u{{\bf u}}\def\e{{\bf e}}\def\btau{\boldsymbol{\tau}}\def\bgam{\boldsymbol{\gamma}}
\def\gamd{\dot\bgam}\def\De{{\rm De}}
\def\n{{\bf n}}\def\E{{\bf E}}\def\T{{\bf T}}
\def\dgamd{\stackrel{\triangledown}{\gamd}}
\def\bs{{\boldsymbol{\sigma}}}
\def\dbtau{\stackrel{\triangledown}{\btau}}\def\x{{\bf x}}\def\U{{\bf U}}
\def\F{{\bf F}}\def\L{{\bf L}}\def\O{{\boldsymbol\Omega}}\def\n{{\bf n}}
\def\BS{{\boldsymbol \Sigma}}\def\I{{\cal I}}
\def\H{{\cal H}}\def\G{{\cal G}}\def\B{{\cal B}}
\def\L{{\bf  L}}\def\M{{\bf M}}\def\N{{\bf N}}\def\H{{\bf H}}
\def\r{{\bf r}} \def\A{{\cal A}}
\def\P{{\bf  P}}\def\Q{{\bf Q}}

\begin{document}

\title{Integral theorems for locomotion in complex fluids}
\title{Locomotion in complex fluids: Integral theorems}
\author{Eric Lauga}
\email{e.lauga@damtp.cam.ac.uk}
\affiliation{Department of Applied Mathematics and Theoretical Physics,  University of Cambridge, Wilberforce Road, Cambridge, CB3 0WA, United Kingdom}
\date{\today}
\begin{abstract}
The biological fluids encountered by self-propelled cells  display complex microstructures and  rheology. We consider here  the general problem of low-Reynolds number locomotion in a complex fluid.  {Building on classical work on the transport of particles in viscoelastic fluids,} we demonstrate how to mathematically derive three integral theorems relating the arbitrary motion of an isolated organism to its swimming kinematics {in a non-Newtonian fluid}. These theorems correspond to three situations of interest, namely (1) squirming motion in a linear viscoelastic fluid, (2) arbitrary surface deformation in a weakly non-Newtonian fluid, and (3) small-amplitude deformation in an arbitrarily non-Newtonian fluid.  Our final results, valid for a wide-class of {swimmer geometry,} surface kinematics and constitutive models, at most require mathematical knowledge of a series of Newtonian flow problems, and will be useful to quantity the locomotion of biological and synthetic swimmers in complex environments. 
\end{abstract}
\maketitle

\section{Introduction}

Among all  active fields of fluid mechanics, the biological hydrodynamics  of cellular life has recently undergone a bit of a renaissance  \cite{braybook}.  This is due to three facts. First, while the hydrodynamics of swimming cells primarily interested scientists  from traditional  continuum mechanics \cite{lighthill75,brennen77,childress81,lp09,stocker}, a number of problems in collective locomotion have found traction in the  condensed matter physics community, with many questions still under active debate \cite{ramaswamy10}. Second,  new quantitative data from  the biological world has led to renewed interest in classical questions, in particular regarding the  synchronization of cellular appendages \cite{lg12}. The third reason, and the one at the center of our study, concerns locomotion in fluids displaying non-Newtonian characteristics. 

In most biological situations, the fluids encountered by self-propelled cells  display complex microstructures and  rheology. Some bacteria progress through multi-layered host tissues while others live in open water surrounded by particle suspensions \cite{Brock}. Lung cilia have to transport viscoelastic, polymeric mucus \cite{sleigh88}. Mammalian spermatozoa have to overcome the resistance of cervical mucus in order to qualify for the race to the finish line \cite{suarez06}. In all these situations, a non-Newtonian fluid is being transported, or being exploited to induce fluid transport, and it is of fundamental importance to quantify the relationship between  kinematics  and the resulting transport.

The problem of predicting the swimming speed of a low-Reynolds swimmer in a complex fluid was first addressed in three pioneering studies  focusing on a  two-fluid model 
\cite{ross1974}, second-order fluid \cite{chaudhury1979}, and linearly viscoelastic fluids \cite{fulford1998}. Recent work started by looking at the asymptotic regime of small-amplitude waving motion in Oldroyd-like fluids  \cite{lauga07,FuPowersWolgemuth2007,FuWolgemuthPowers2009}, predicting that, for a fixed swimming gait,  the swimming speed is always smaller than in a Newtonian fluid.  
Importantly, that result does not appear to depend on the detail of the continuum description for the viscoelastic fluid, and is unchanged for more advanced  nonlinear relationships such as FENE (finitely extensible nonlinear elastic) or Giesekus models in the same asymptotic limit \cite{lauga07}. Numerical computations in two dimensions were then employed to probe the limit of validity of these results. While they confirmed the low-amplitude results, they also demonstrated that for some large-amplitude motion  viscoelasticity could actually enhance the swimming speed of the model cell \cite{teran2010}. In contrast, simulations for spherical squirmers --  swimmers acting on the surrounding fluid  tangentially to their shape -- showed that viscoelastic swimming was systematically slower than its Newtonian counterpart even at high Weissenberg number  \cite{lailai-pre,laipof1}. 

Beyond polymeric fluids,  analytical modeling was also proposed for locomotion in fluids displaying other rheological behavior. The two-dimensional approach was applied to swimming in a gel  \cite{fu10}, a two-phase fluid  \cite{du12}, and  yield stress materials  \cite{balmforth10}. A series of models was exploited to demonstrate that locomotion in a heterogeneous media -- one made of stationary rigid inclusions -- could systematically enhance self-propulsion  \cite{leshansky09}. Inelastic fluids with shear-dependent viscosities were also considered. While they necessarily impact the fluid motion at a higher order than polymeric stresses \cite{rodrigo2013}, it was shown that shear and therefore rheological  gradients  along the swimmer could lead to swimming enhancement  \cite{montenegro12,montenegro13}.  Different setups were also proposed and tested to demonstrate that  nonlinearities in the fluid rheology could be exploited to design novel actuation and swimming devices   
  \cite{normand08,lauga_life,pak10,pak12,keim12}.

In contrast with theoretical studies, detailed experimental work on the fluid mechanics of swimming in complex fluids has been limited to a small number of investigations. A study of the nematode {\it C.~elegans} self-propelling  in synthetic polymeric solutions behaving  as Boger fluids (constant shear viscosities) showed a systematic decrease of their swimming speed \cite{arratia2011} consistent with asymptotic theoretical predictions   \cite{lauga07,FuPowersWolgemuth2007}.  In contrast, recent work on a two-dimensional rotational model of a swimming sheet demonstrated that  Boger fluids always lead to an increase of the swimming speed   while  elastic fluids with shear-thinning viscosities lead to a systematic decrease  \cite{dasgupta13}.  The swimming increase in Boger fluid was also obtained  in the case of force-free flexible swimmers driven by oscillating magnetic fields   \cite{espinosa13}. Translating rigid helices used as a model for  free-swimming of bacteria were further shown to also decrease their swimming speed at small helix amplitude but displayed a modest speed increase for larger  helical amplitude   \cite{Liu2011}. This increase is consistent with earlier computations  \cite{teran2010} and was further confirmed by a detailed  numerical study \cite{spagnolie2013}. 

In this paper, we  consider theoretically the general problem of low-Reynolds number locomotion in a non-Newtonian fluid. Following classical work proposing   integral formulations to quantify cell locomotion in Newtonian flows \cite{stone96} and the motion of solid particles in viscoelastic fluids {\cite{leal75,ho76,brunn76a,brunn76b,leal79,leal80} (themselves adapted from earlier work on inertial effects \cite{brenner63,cox65,cox68, ho74})},  
we  demonstrate how to mathematically derive three integral theorems relating the arbitrary motion of an organism to its swimming kinematics. 
After introducing the mathematical setup (\S\ref{sec:setup}) and recalling the classical results for locomotion in a Newtonian fluid (\S\ref{sec:Newtonian}), the first theorem considers the  classical tangential squirmer model of Lighthill and Blake (\S\ref{sec:LVFS}). We demonstrate that in this case, in an arbitrary linear viscoelastic fluid the swimming kinematics are the same as in a Newtonian fluid. The second theorem considers the asymptotic limit of small deviation from the Newtonian behavior (low Deborah number limit) with no asymptotic constraint on the amplitude of the deformation (\S\ref{sec:weakly}). We compute analytically in this weakly non-Newtonian regime the first-order effect of the non-Newtonian stresses on the swimming kinematics. In the final, {and more general},  theorem we address an arbitrary nonlinear viscoelastic fluid and  derive the  swimming kinematics in the limit of small-amplitude deformation  (\S\ref{sec:small}). The theorems in \S\ref{sec:weakly} and \S\ref{sec:small} address therefore two complementary asymptotic limits: small deformation rate in \S\ref{sec:weakly}  (low Deborah and Weissenberg numbers) 
vs.~small deformation amplitude in \S\ref{sec:small} 
(low Weissenberg, arbitrary Deborah). The implications of our results for Purcell's scallop theorem are then discussed in \S\ref{sec:scallop}. Finally, we apply the general theorem from \S\ref{sec:small}  to the locomotion of a sphere in an Oldroyd-B fluid in \S\ref{sec:Old}. We show in particular that we can construct swimming kinematics which are either enhanced or reduced by the presence of viscoelastic stresses, thereby further demonstrating that the impact of non-Newtonian rheology on swimming is kinematics-dependent.



\section{Mathematical setup}
\label{sec:setup}

\begin{figure}[t]
\begin{center}
 \includegraphics[width=0.4\textwidth]{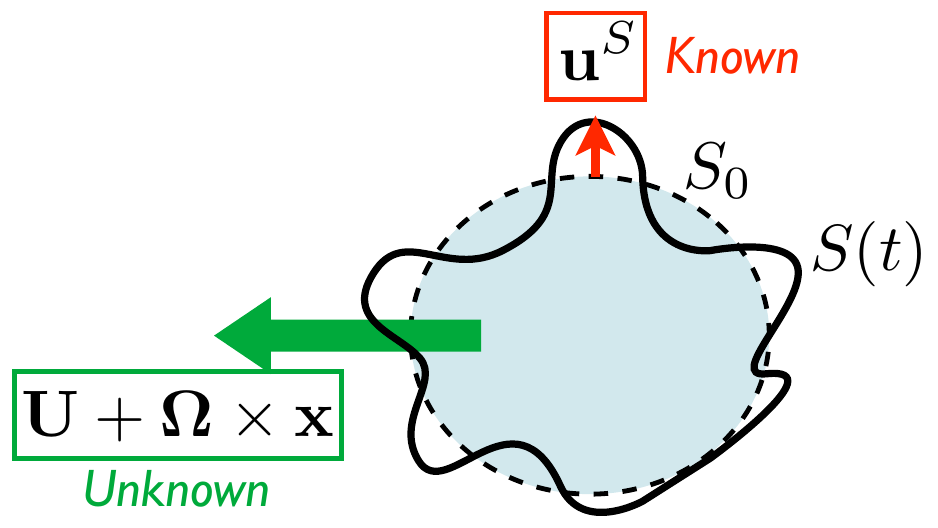}
\end{center}
 \caption{Schematic representation of the  swimming problem: Material points on a  surface $S_0$ are moving periodically to a time-dependent shape $S(t)$. The instantaneous velocity on the surface is denoted   $\u^S$, and is the swimming gait.  As a result of free swimming motion, the shape $S(t)$ moves instantaneously with three-dimensional solid body velocity  $\U(t)$ and rotation rate $\O(t)$.}
\label{general}
\end{figure}

The mathematical setup for the swimming problem is illustrated in Fig.~\ref{general}. We consider a closed surface $S_0$ undergoing periodic  deformation into a shape denoted $S(t)$. This shape is that of an isolated three-dimensional swimmer self-propelling in an infinite fluid. 
We use the notation $\x^S$ for the instantaneous location of the material points on the surface of the swimmer and $\n$ the instantaneous normal to the surface $S(t)$.  The velocity field and stress tensor in the fluid are written $\u$ and  $\bs$  respectively. The stress is given by $\bs=-p{\bf 1} + \btau$ where $p$ is the pressure, $\bf 1$ the identity tensor, and $\btau$ the deviatoric stress, modeled by specific constitutive relationships considered in the following sections.  The equations to solve for the fluid are the incompressibility condition,  $\nabla\cdot \u =0$, and Cauchy's equation of motion in the absence of inertia 
\begin{equation}\label{Cauchy}
\nabla p =  \nabla \cdot  \btau.
\end{equation}
The boundary conditions for  Eq.~\eqref{Cauchy} are given by 
\begin{equation}\label{BC_general}
\u (\x^S,t) =   \U + \O\times \x^S + \u^S,
\end{equation}
where the imposed surface velocity,  $\u^S(\x^S,t)$, is the swimming gait, and $\{\U,\O\}$ are the unknown swimming kinematics, i.e.~the instantaneous solid body translation and rotation of the shape $S(t)$. Both are to be determined by enforcing the {instantaneous} condition of no net force or torque on the swimmer as
\begin{equation}\label{noF}
\int\!\!\!\int _{S(t)}\bs\cdot\n\,\d S=\int\!\!\!\int _{S(t)}\x^S\times (\bs\cdot\n) \, \d S={\bf 0}
\end{equation}
Note that throughout the paper we will use the notation  $\gamd =\nabla \u +\!^t\nabla\u $  for the shear rate tensor, $\gamd$, equal to twice the symmetric rate-of-strain tensor {($^t$ denotes the transpose of a tensor)}. Note also that surface motion ($\u^S\neq \bf 0$) does not necessarily imply a change in shape as only the components of $\u^S$ normal to the surface, $\u^S\cdot \bf n$, contribute to the deformation of the shape.

\section{Newtonian case}
\label{sec:Newtonian}

Before addressing the non-Newtonian case, we briefly summarize here the integral theorem in the Newtonian case for which $\btau=\mu \gamd$. This is work originally presented by  Stone and Samuel \cite{stone96} based on an application of Lorentz' reciprocal theorem. 

We consider two solutions of  Stokes flow with the same viscosity  around the instantaneous surface $S(t)$. The first one has velocity and stress fields given by $(\u,\bs)$ and is that of the swimming problem. Its boundary conditions are thus yet to be determined. The second solution, denoted $(\hat\u,\hat\bs)$, is 
the problem of solid body motion with instantaneous shape $S(t)$, with force $\hat\F$, 
\begin{equation}\label{force}
\hat\F=\int\!\!\!\int \hat\bs \cdot \n\,\d S,
\end{equation}
and torque  $\hat\L$ with respect to some origin in the body, 
\begin{equation}\label{torque}
\hat\L=\int\!\!\!\int \x^S \times (\hat\bs \cdot \n)\,\d S.
\end{equation}
In the hat problem, the shape $S(t)$ moves thus instantaneously like a solid body with   with velocity $\hat\U$ and rotation speed $\hat\O$, and thus on the surface we have
\begin{equation}\label{hatu_SB}
\hat\u  =  \hat \U + \hat\O\times \x^S,
\end{equation}
for all material points $ \x^S$.

In the absence of body forces, Lorentz' reciprocal theorem states if both problems concern a fluid with identical viscosity we have the equality of virtual powers 
\begin{equation}\label{reciprocal}
\int\!\!\!\int _{S}\u\cdot \hat\bs\cdot \n \,\d S=
\int\!\!\!\int _{S}\hat\u\cdot \bs\cdot \n \,\d S.
\end{equation}
Since $\hat\u$ is known everywhere on the surface, Eq.~\eqref{hatu_SB}, the 
 left term in Eq.~\eqref{reciprocal} gives
\begin{equation}
\int\!\!\!\int _{S}\hat\u\cdot \bs\cdot \n \,\d S = \hat\U\cdot
\int\!\!\!\int _{S}\bs\cdot\n\,\d S
+\hat\O\cdot
\int\!\!\!\int _{S}\x^S\times (\bs\cdot\n) \, \d S=0,
\end{equation}
because swimming is force- and torque-free at all instants, see Eq.~\eqref{noF}. Consequently, Eq.~\eqref{reciprocal} simplifies to
\begin{equation}\label{9}
\int\!\!\!\int _{S}\u\cdot \hat\bs\cdot \n \,\d S=0.
\end{equation}
By using the kinematic decomposition on the swimmer surface in Eq.~\eqref{BC_general}, Eq.~\eqref{9} becomes 
\begin{equation}
\int\!\!\!\int _{S}\u\cdot \hat \bs\cdot \n \,\d S = \U\cdot
\int\!\!\!\int _{S}\hat \bs\cdot\n\,\d S
+\O\cdot
\int\!\!\!\int _{S}\x^S\times (\hat\bs\cdot\n) \, \d S
+\int\!\!\!\int _{S}\u^S\cdot \hat \bs\cdot \n \,\d S
=0
\end{equation}
and thus, using Eqs.~\eqref{force} and \eqref{torque} we finally obtain
\begin{equation}\label{N_final}
\hat \F\cdot \U +\hat \L \cdot \O = -\int\!\!\!\int _{S}\u^S\cdot\hat \bs \cdot \n \,\d S.
\end{equation}

The final result, Eq.~\eqref{N_final}, is an equation for the swimming kinematics, $\{\U,\O\}$. In order to solve that equation, one needs to know the  distribution of stress,  $\hat \bs \cdot \n$, on the surface $S$ for solid body motion in a Newtonian flow under and external force $\hat \F$ and torque $\hat \L$, which we assume is known. Since the values of $\hat \F$  and $\hat \L$ are arbitrary, Eq.~\eqref{N_final} allows us to solve for all components of $\U$ and $\O$. 

As a side note which will be exploited later in the paper, we remind that the two velocity and stress fields in the application of Lorentz' reciprocal theorem correspond to two problems in the same Newtonian fluid. However, this constraint is relaxed in the final result quantified by Eq.~\eqref{N_final}. This is because the left-hand side of Eq.~\eqref{reciprocal} turns out to be identically zero and a solid body motion implies zero virtual rate of work against a distribution of stress from force-free and torque-free swimming. Another way to see this is to note that by changing  the fluid viscosity in  Eq.~\eqref{N_final},   both sides of the equation are modified by the same prefactor since  forces, torque, and stresses all scale proportionally with the viscosity in the Stokes regime.

\section{Squirming in a linearly viscoelastic fluid}
\label{sec:LVFS}

\subsection{Squirming}
In this section we present the derivation for the first of our integral theorems. We consider here the class of swimmers known  as squirmers which {deform their surfaces everywhere} in the direction parallel to their shapes, i.e.~for which $\u^S\cdot\n=0$ everywhere and for all times. The shape of the swimmer is therefore  fixed in time, $S_0$, and the distribution of velocity $\u^S$ is assumed to be known on $S_0$ ($\u^S$ does not have to be steady, as we see below). This squirmer model, most often used when  $S_0$ is a sphere, was first proposed by Lighthill \cite{Lighthill1952}, with corrections by Blake \cite{Blake1971a}, and is one of the very few  analytical solutions to low-Reynolds swimming. As such, it has proven very popular to address a larger number of fundamental problems in cell locomotion, including hydrodynamic interactions \cite{Ishikawa2006},  the  rheology of swimmer suspensions \cite{Ishikawa2007}, optimal locomotion \cite{Michelin2010}, nutrient uptake \cite{Magar2003,Michelin2011},  inertial swimming  \cite{Wang2012} and locomotion in polymeric fluids 
\cite{lailai-pre,laipof1}.

\subsection{Generalized linear viscoelastic fluid}
For the constitutive relationship, we assume in this first section  that the fluid is a generalized linear viscoelastic fluid \cite{birdvol1,morrison}. {Admittedly, this is a very idealized assumption as the flow around a swimming cell is non-viscometric while the linear constitutive equation only applies to small-amplitude viscometric motions. However, within this idealized class of fluids, we are able to obtain the solution for the swimming problem exactly without requiring  any asymptotic expansion, which makes it valuable as an academic exercise.  Furthermore, the work in this section will in fact represent the leading-order behavior for a fluid with a more complex, nonlinear rheology as addressed in \S\ref{sec:small} asymptotically, and therefore the mathematical details outlined below are important.}

A generalized linear viscoelastic fluid is characterized by  arbitrary relaxation modulus, $G$, such that the stress is linearly related to the history of the rate of train in the most general form as
\begin{equation}\label{LVE}
\btau(\x,t)=\int_{-\infty}^t G(t-t') \gamd (\x,t')\,\d t',
\end{equation}
or, using index notation, 
\begin{equation}\label{LVE_tensor}
\tau_{ij}(\x,t)=\int_{-\infty}^t G(t-t') \dot\gamma_{ij} (\x,t')\,\d t'.
\end{equation}

In order to derive the integral theorem in this section we are going to use Eq.~\eqref{LVE_tensor} written in Fourier space. This will allow us to derive an integral theorem for each Fourier components of the swimming kinematics {(see earlier work on the so-called correspondence principle for linear viscoelasticity \cite{graham73}}).  
The one-dimensional Fourier transform and its inverse are defined for any function $f(t)$  as
\begin{equation}\label{def_F}
\tilde f (\omega)=\frac{1}{\sqrt{2\pi}} \int_{-\infty}^\infty f(t) e^{-i \omega t} \,\d t,\quad
f(t)=\frac{1}{\sqrt{2\pi}} \int_{-\infty}^\infty \tilde f(\omega) e^{i \omega t} \,\d t.
\end{equation}
{Following a classical textbook approach \cite{morrison}, we apply} the Fourier transform to Eq.~\eqref{LVE_tensor},  leading to 
\begin{equation}
\tilde \tau_{ij} (\x,\omega)
=\frac{1}{\sqrt{2\pi}} \int_{-\infty}^\infty \tau_{ij} (\x,t) e^{-i \omega t} \,\d t
=\frac{1}{\sqrt{2\pi}} \int_{-\infty}^\infty\left[ \int_{-\infty}^t G(t-t') \dot\gamma_{ij} (\x,t')\,\d t' \right]e^{-i \omega t} \,\d t.
\end{equation}
Change the order of time-integration allows us to obtain
\begin{equation}
\tilde \tau_{ij} (\x,\omega)
=\frac{1}{\sqrt{2\pi}} \int_{-\infty}^\infty\left[ \int_{t'}^{\infty} G(t-t') e^{-i \omega t}\,\d t \right]  \dot\gamma_{ij} (\x,t')\,\d t'.
\end{equation}
We then write $e^{-i \omega t}=e^{-i \omega (t-t')}e^{-i \omega t'}$ and get
\begin{equation}
\tilde \tau_{ij} (\x,\omega)
=\frac{1}{\sqrt{2\pi}} \int_{-\infty}^\infty\left[ \int_{t'}^{\infty} G(t-t') 
e^{-i \omega (t-t')}
\,\d t \right]  \dot\gamma_{ij} (\x,t') e^{-i \omega t'}\,\d t'.
\end{equation}
A final  change of variable $\bar t=t-t'$ in the bracketed  integral leads to 
\begin{equation}\label{almost}
\tilde \tau_{ij} (\x,\omega)
=\frac{1}{\sqrt{2\pi}} \int_{-\infty}^\infty\left[ 
\int_{0}^{\infty} G(\bar t) 
e^{-i \omega \bar t}
\,\d \bar t
 \right]  \dot\gamma_{ij} (\x,t') e^{-i \omega t'}\,\d t'.
\end{equation}
Defining
\begin{equation}\label{19}
\G(\omega)=\int_{0}^{\infty} G(\bar t) 
e^{-i \omega \bar t}
\,\d \bar t,
\end{equation}
we are able to take $\G(\omega)$ out of the integral relationship in Eq.~\eqref{almost}, leading to
\begin{equation}\label{CR_F}
\tilde \tau_{ij} (\x,\omega)
=
\G(\omega)\tilde {\dot\gamma}_{ij} (\x,\omega).
\end{equation}
The statement in Eq.~\eqref{CR_F} is the constitutive relationship written in Fourier space, {while Eq.~\eqref{19} is the classical approach to relate  the relaxation modulus of the fluid to the storage and loss modulus   in Fourier space \cite{morrison}.}

\subsection{Integral theorem}
In order to derive the integral theorem, we first rewrite the swimming problem in Fourier space. Since the kinematics is restricted to squirming motion, the shape of the swimmer does not change, and $\u^S$ is known with no ambiguity in the Eulerian frame for each  point $\x^S$ and for all times. We therefore decompose the surface velocity in Fourier modes as
\begin{equation}
\u^S(\x^S,t)=\frac{1}{\sqrt{2\pi}} \int_{-\infty}^\infty \tilde \u^S(\x^S,\omega) e^{i \omega t} \,\d t,
\end{equation}
and do similarly for  the swimming kinematics as
\begin{equation}
\{\U(t),\O(t)\}=\frac{1}{\sqrt{2\pi}} \int_{-\infty}^\infty \{\tilde\U(\omega),\tilde\O(\omega)\} e^{i \omega t} \,\d t.
\end{equation}
The Fourier transforms of the velocity and pressure fields are similarly defined.  

Using Eq.~\eqref{CR_F}, we then see that the incompressible Cauchy's equation, Eq.~\eqref{Cauchy},  becomes  in Fourier space
\begin{equation}\label{CauchyF}
\nabla  \tilde p (\x,\omega) =\G (\omega) \nabla^2 \tilde \u (\x,\omega)
,\quad 
\nabla\cdot \tilde \u  (\x,\omega) = 0. 
\end{equation}
Consequently, the  swimming problem consists in solving Eq.~\eqref{CauchyF}
with the boundary condition
\begin{equation}\label{BCF}
\tilde\u(\x^S,\omega)
= \tilde\U(\omega) + \tilde\O(\omega) \times \x^S + \tilde\u^S(\x^S,\omega).
\end{equation}
The problem defined by Eqs.~\eqref{CauchyF}-\eqref{BCF} is a Stokes flow locomotion problem with (complex) viscosity $\G (\omega)$. The integral theorem of \S\ref{sec:Newtonian} is then directly applicable, and we have
\begin{equation}\label{swim_Fmode}
\hat \F\cdot \tilde\U(\omega) +\hat \L \cdot \tilde \O (\omega)
= -\int\!\!\!\int _{S}\n\cdot\hat \bs \cdot \tilde\u^S(\x^S,\omega)\,\d S.
\end{equation}

The final step allowing us to go back from Fourier to real space is to take advantage  of the fact that  the hat problem in Eq.~\eqref{swim_Fmode} is a Newtonian Stokes flow with arbitrary viscosity (see the discussion at the end of \S\ref{sec:Newtonian}).  We can take it to be a constant reference viscosity, $\mu_0$, independent of the frequency $\omega$. Furthermore, the shape $S$ of the swimmer is not a function of time. We therefore see that none of the terms in 
Eq.~\eqref{swim_Fmode} depend on the frequency except for the three Fourier components: $\U(\omega)$, 
$\O(\omega)$, and $\tilde\u^S(\x^S,\omega)$. 
The inverse Fourier transform in Eq.~\eqref{def_F} can directly be applied to Eq.~\eqref{swim_Fmode}  leading to  the same integral equation as for the Newtonian case
\begin{equation}\label{integral_1}
\hat \F\cdot \U +\hat \L \cdot \O = -\int\!\!\!\int _{S}\n\cdot\hat \bs \cdot \u^S\,\d S.
\end{equation}

In summary, for squirming in an arbitrary linear viscoelastic fluid we obtain an exact integral theorem for the swimming kinematics, Eq.~\eqref{integral_1}, identical to the Newtonian one. The  squirming velocity and rotation rate in a linearly viscoelastic fluid are thus  identical to those in a Newtonian fluid.  In Eq.~\eqref{integral_1} the hat problem is in a different fluid though, namely a Newtonian Stokes flow with constant, arbitrary,  viscosity. It is notable that no asymptotic assumption was required to derive  Eq.~\eqref{integral_1}.

Two assumptions   were necessary in order to derive this result. First we assumed that the motion was always tangential to the shape, allowing us to write the boundary condition on the swimmer surface in Fourier space  and to take the inverse Fourier transform of Eq.~\eqref{swim_Fmode} with no ambiguity.
 Second, we assumed that the fluid was linearly viscoelastic with no nonlinear rheological response ({despite the shortcomings of this assumption, as outlined above}). Beyond this, no restriction was required on the distribution of surface velocity,  $\u^S$, and in particular it could be unsteady. If either assumption breaks down, and the fluid is nonlinear (as most fluids are) or the swimmer undergo normal shape deformation, an asymptotic analysis will be required, as we show in the  following sections. 

\section{Swimming in weakly non-Newtonian fluids}
\label{sec:weakly}

\subsection{Weakly non-Newtonian rheology}

In this second section we consider fluids whose rheological behaviors are close to that of a Newtonian fluid. If a fluid displays a zero-shear-rate Newtonian behavior, then we are concerned here in situations in which the fluid is deformed at small shear rate, and we will quantify the first effect of non-Newtonian rheology.  

Two specific examples of  such fluids can be given. For an inelastic fluid with shear-dependent viscosity $\eta$ (so-called Generalized Newtonian fluids), we are interested in the limit where  $({\eta-\eta_0})/{\eta_0}\ll 1$ when $\eta_0$ is the zero-shear-rate viscosity \cite{morrison}. An another example is that of elastic fluids at small Deborah numbers, $\De \ll 1$, for which the constitutive relationship is the retarded motion expansion \cite{birdvol1}.

In all cases, we assume that the non-Newtonian rheology of the fluid is a small perturbation, of dimensionless size $\epsilon$, on an otherwise Newtonian dynamics. We thus write the constitutive relationship in the most general form  as 
\begin{equation}
\btau = \eta \gamd + \epsilon \BS [\u],
\end{equation}
where  $ \BS [\u]$ is {a symmetric tensor} and an arbitrary nonlinear functional of $\u$ with units of stress and $\epsilon\ll 1$ quantifies the small deviation from Newtonian behavior. For example, $\epsilon$ could be a small Deborah number in the case of viscoelastic fluids, or a small Carreau number for a shear-thinning fluid. Importantly, since we assume a small value for $\epsilon$ we have no time-history in the constitutive relationship and therefore the shape $S(t)$ will be allowed to vary arbitrarily in time.

\subsection{Integral theorem}
{In order to derive the integral theorem in this case, we  adapt below classical work on the first effect of non-Newtonian rheology on the dynamics of small particles in externally-driven flows (see e.g.~classical studies in Refs.~\cite{leal75,ho76,brunn76a,brunn76b} and reviews in Refs.~\cite{leal79,leal80}) to the case of self-propulsion.  The reader familiar already familiar with these works will not be surprised by the expected form of the non-Newtonian component of the swimming speed derived  in Eq.~\eqref{integral_2}.
}

\subsubsection{Asymptotic expansion}

We look for 
regular perturbation expansions for all variables under the form 
\begin{equation}
\{\mathbf{u},\btau,p,\bs\}= 
\{\mathbf{u}_{0} ,{\btau}_{0},p_0, \bs_0 \} 
+ \epsilon\{\mathbf{u}_{1},{\btau}_{1},p_1, \bs_1 \}+...
\end{equation}
and similarly for the resulting locomotion kinematics
\begin{equation}
\{\U,\O\}= 
\{\U_0,\O_0\}
+ \epsilon\{\U_1,\O_1\}+...,
\end{equation}
{which, in the most general case,  are allowed to depend in time}.

The swimming gait, $\u^S$, is imposed at order $\epsilon^0$ and has no component at higher orders. In other words the swimming gait is fixed and independent of the rheological behavior of the fluid. On  the swimmer surface we thus have the {instantaneous} boundary conditions at order 
$\epsilon^0$ and $\epsilon$ given by
\begin{subeqnarray}
\u_0 & = &  \U_0 + \O_0\times \x + \u^S,\\
\u_1 & = &  \U_1 + \O_1\times \x.\slabel{BC_order1}
\end{subeqnarray}

The hydrodynamic force and torque on the swimmer are given by 
\begin{equation}
\F(t)=\int\!\!\!\int _{S(t)}\n\cdot\bs \,\d S,\quad
\L(t)=\int\!\!\!\int _{S(t)}\x^S\times (\bs\cdot\n) \,\d S,
\end{equation}
where the torque can be computed with respect to an arbitrary origin since $\F=\bf 0$. Expanding both in powers of $\epsilon$ we obtain
\begin{equation}
\{\F,\L\}=\{\F_0,\L_0\}+\epsilon\{\F_1,\O_1\}+...,
\end{equation}
and we see that the force- and torque-free requirements leads to $\F_i=\O_i=\bf 0$ at any order $i$ {for all times}.

\subsubsection{Order $\epsilon^0$}

At order $\epsilon^0$, the flow is Newtonian, $\bs_0 = -p_0{\bf 1} +\eta \gamd_0$,   and we can directly apply the  integral result from \S\ref{sec:Newtonian}
\begin{equation}
\hat \F\cdot \U_0 +\hat \L \cdot \O_0 = -\int\!\!\!\int _{S(t)}\n\cdot\hat \bs \cdot \u^S\,\d S,
\end{equation}
where $S(t)$ is the instantaneous shape of the swimmer 
(note that we placed no restriction on the amplitude of the surface motion).

\subsubsection{Order $\epsilon$}

At next order, we are interested in  deriving the new formulae leading to $\U_1$ and $\O_1$. At order $\epsilon$, the constitutive relationship is written as
\begin{equation}
\bs_1 = -p_1{\bf 1} +\eta \gamd_1 + \BS[\u_0].
\end{equation}

In order to derive the integral result, we first have to use a modified version of LorentzÕ reciprocal theorem. We start by noting that we have, {at each instant},
\begin{equation}\label{start}
\nabla\cdot \bs_1 = 0 = \nabla\cdot \hat \bs,
\end{equation}
where the hat stress field, $\hat \bs$, refers to the Stokes flow where the body is subject to external force, $\hat\F$,  and an external torque, $\hat\O$, in Newtonian fluid of viscosity $\eta$ (same notation as in \S\ref{sec:Newtonian}).  We then dot Eq.~\eqref{start} with the velocity fields $\hat\u$ and $\u_1$ as 
\begin{equation}\label{virtualW}
\hat\u\cdot \nabla\cdot \bs_1 =\u_1\cdot \nabla\cdot \hat \bs,
\end{equation}
which states that the virtual rates of working of each flow  in the opposite stress field are equal. Integrating 
Eq.~\eqref{virtualW} 
 over the entire fluid volume, $V(t)$, and using the divergence theorem leads to the equality
\begin{equation}\label{4terms}
\int\!\!\!\int _{S(t)}\n\cdot \hat\bs\cdot \u_1\,\d S
-
\int\!\!\!\int _{S(t)}\n\cdot \bs_1\cdot \hat\u\,\d S
=
\int\!\!\!\int\!\!\!\int_{V(t)} \bs_1:\nabla \hat\u \, \d V
-
\int\!\!\!\int\!\!\!\int_{V(t)} \hat\bs:\nabla \u_1 \, \d V
,\end{equation}
where the normal $\n$ is directed into the fluid.
Examining the right-hand side of Eq.~\eqref{4terms} we can rewrite it as
\begin{align}
\nonumber \int\!\!\!\int\!\!\!\int_{V(t)} \bs_1:\nabla \hat\u \, \d V
-
\int\!\!\!\int\!\!\!\int_{V(t)} \hat\bs:\nabla \u_1 \, \d V
=
\int\!\!\!\int\!\!\!\int_{V(t)} \BS[\u_0] :\nabla \hat\u \, \d V\\
+ 
\int\!\!\!\int\!\!\!\int_{V(t)}
\left[
(-p_1{\bf 1} + \eta\gamd_1):\nabla\hat\u
-
(-\hat p{\bf 1} + \eta \hat\gamd):\nabla\u_1
\right] \, \d V.
\end{align}
Using incompressibility for the fields $\u_1$  and $\hat\u$ (i.e.~$\nabla\cdot \u_1=\nabla\cdot \hat\u=0$), it is straightforward to show that
\begin{equation}\label{goestozero}
\int\!\!\!\int\!\!\!\int_{V(t)}
\{
(-p_1{\bf 1} + \eta \gamd_1):\nabla\hat\u
-
(-\hat p{\bf 1} +\eta  \hat\gamd):\nabla\u_1
\} \, \d V=
\int\!\!\!\int\!\!\!\int_{V(t)} \eta
\{
 \gamd_1:\nabla\hat\u
- \hat\gamd:\nabla\u_1
\} \, \d V,
\end{equation}
which is zero by symmetry, so that Eq.~\eqref{4terms} becomes
\begin{equation}\label{3terms}
\int\!\!\!\int _{S(t)}\n\cdot \hat\bs\cdot \u_1\,\d S
-
\int\!\!\!\int _{S(t)}\n\cdot \bs_1\cdot \hat\u\,\d S
=
\int\!\!\!\int\!\!\!\int_{V(t)} \BS[\u_0] :\nabla \hat\u \, \d V.
\end{equation}
In the hat problem, the surface instantaneously moves with  solid-body motion with velocity $\hat\U$ and rotational speed $\hat\O$, and therefore  on the surface of the swimmer, we have $\hat\u = \hat\U + \hat\O\times \x^S$ Consequently,  the second integral on the left-hand-side of Eq.~\eqref{4terms} is given by
\begin{equation}\label{F1}
\int\!\!\!\int _{S(t)}\n\cdot \bs_1\cdot \hat\u\,\d S=
\hat\U\cdot \int\!\!\!\int _{S(t)}\n\cdot \bs_1 \,\d S 
+\hat\O\cdot 
\int\!\!\!\int _{S(t)}\x^S \times( \n\cdot \bs_1)\,\d S.
\end{equation}
The two integrals on the right-hand-side of 
Eq.~\eqref{F1} are the  {instantaneous} first-order force and torque on the swimmer, which, as was shown above,  are both zero  and thus we obtain 
\begin{equation}
\int\!\!\!\int _{S(t)}\n\cdot \bs_1\cdot \hat\u\,\d S=0.
\end{equation}
As a consequence,  Eq.~\eqref{3terms} simplifies to
\begin{equation}\label{2terms}
\int\!\!\!\int _{S(t)}\n\cdot \hat\bs\cdot \u_1\,\d S
=
\int\!\!\!\int\!\!\!\int_{V(t)} \BS[\u_0] :\nabla \hat\u \, \d V.
\end{equation}
On the swimmer surface, we then apply the boundary condition at order $\epsilon$ from Eq.~\eqref{BC_order1}  and obtain the final integral relationship
\begin{equation}\label{integral_2}
\hat \F\cdot \U_1 +\hat \L \cdot \O_1 = \int\!\!\!\int\!\!\!\int_{V(t)} \BS[\u_0] :\nabla \hat\u \, \d V.
\end{equation}

This second integral theorem, Eq.~\eqref{integral_2}, allows us to compute the first non-Newtonian correction to the the Newtonian swimming kinematics, namely $\U_1$ and 
$\O_1$, using only the knowledge from Newtonian solution. {Importantly, the derivation is instantaneous, and it is thus  valid for both steady and unsteady problems}.  
In contrast to the Newtonian integral theorem, we notice that we need to know more than just the solution to the hat problem and the entire velocity field, $\u_0$, for the {instantaneous} Newtonian swimming problem also needs to be known. Given Eq.~\eqref{N_final}, we know the boundary conditions for $\u_0$ and thus solving for it is the same level of complexity as solving for $\hat\u$. 
With the knowledge of both $\u_0$ and $\hat\u$, the volume integral  on the right-hand-side of Eq.~\eqref{integral_2} can be computed, giving access to the swimming kinematics. As a side note, it is clear that the antisymmetric part of $\nabla\hat \u$ does not contribute to Eq.~\eqref{integral_2} since $\BS$ is a symmetric tensor, and thus the integral theorem can also be rewritten as
\begin{equation}\label{integral_2_bis}
\hat \F\cdot \U_1 +\hat \L \cdot \O_1 = \int\!\!\!\int\!\!\!\int_{V(t)} \BS[\u_0] : 
\hat\e \, \d V,
\end{equation}
where $\hat\e =\frac{1}{2}(^t\nabla\hat \u+\nabla\hat \u )$ is the symmetric rate-of-strain tensor for the hat problem. 

\section{Small-amplitude swimming in nonlinear fluids}
\label{sec:small}

For the  two integral theorems above  we considered very specific constitutive relationships. Specifically, in order to derive Eq.~\eqref{integral_1} we assumed that the fluid rheology was linear while, in order to obtain Eq.~\eqref{integral_2}, we allowed some nonlinearity in the constitutive relationship but assumed it was always small. It would be desirable to have a theorem valid when the rate of deformation of the fluid is comparable to its relation time, thereby displaying possible nontrivial nonlinear effects on the swimming  kinematics. In order to allow finite values of the Deborah number while deriving the result analytically we consider  another asymptotic limit, namely that of small-amplitude deformations.  {The results presented below are the most important results of this paper and are broadly applicable to different fluids and geometry}. An earlier form of the theorem focusing solely on time-averaged motion was presented in Ref.~\cite{lauga_life}.  {Furthermore, as we detail below,  the results from \S\ref{sec:LVFS} will be used at leading order.}

\subsection{Domain perturbation}

The tool used to derive the approximate solution in this case is that of domain perturbation, as originally proposed by Taylor in his pioneering study of the two-dimensional swimming sheet swimming in a Newtonian fluid \cite{taylor51}. We now denote by $\epsilon$ the dimensionless amplitude of the surface deformation and are interested in deriving the results asymptotically in the limit  $\epsilon\ll 1$. 

In this domain-perturbation approach we have to make explicit the link between the Lagrangian deformation of the surface and the resulting  Eulerian boundary conditions for the solution to the fluid dynamics problem. The reference surface, $S_0$, is described by the field   $\x^S_0$, and we then write the Lagrangian location of material points, $\x^S$,  on the surface as  
\begin{equation}\label{xS}
\x^S(t) = \x^S_0 + \epsilon \x^S_1 (\x^S_0 ,t),
\end{equation}
where $\x^S_1$ represents thus the dimensional change in position  of each reference point  $\x^S_0$. While $\n$ denotes the normal to the surface $S$ into the fluid, we denote by $\n_0$ the normal to the reference surface $S_0$.

We then proceed to solve the problem as a perturbation expansion in powers of $\epsilon$. At order 
 $\epsilon^0$ there is no motion, so we have to go to order $\epsilon$ to obtain the leading-order fluid motion as well as $\epsilon^2$ since we   expect  the swimming kinematics to scale quadratically with the amplitude of the surface motion \cite{taylor51}. We thus write the swimming kinematics as 
 \begin{equation}\label{exp_UO}
\{\U,\O\}= \epsilon\{\U_1,\O_1\}+ \epsilon^2\{\U_2,\O_2\}+...
\end{equation}
and look similarly for velocity and stress fields as 
  \begin{equation}\label{exp}
\{\mathbf{u},\btau,p,\bs\}= \epsilon\{\mathbf{u}_{1} ,{\btau}_{1},p_1, \bs_1 \} 
+ \epsilon^2\{\mathbf{u}_{2},{\btau}_{2},p_2, \bs_2 \}+...,
\end{equation}
which are fields are defined, in the domain-perturbation framework, with boundary conditions on the zeroth-order surface $S_0$. {Note that the 
domain-perturbation approach does rigorously take into account all terms of the dynamics balance for the swimmer, even nonlinear interactions all all orders, as shown in \S\ref{secondo}.}

\subsection{Boundary conditions}
In order to derive the correct boundary conditions for the velocity field in Eq.~\eqref{exp}, we have to pay attention to the kinematics of the surface. The instantaneous boundary condition on the surface of the swimmer is given by
\begin{equation}\label{BCtodo}
\u (\x^S,t) =   \U + \O\times \x^S + \u^S,
\end{equation}
an equation in which all four terms need to be properly expanded in powers of $\epsilon$. The swimming velocity, $\U$, and rotation rate, $\O$, are expanded in Eq.~\eqref{exp_UO} while the expansion for the surface shape is given in Eq.~\eqref{xS}. The  expansion for the swimming gait, $\u^S$,  is carried out using a Taylor expansion on the swimmer surface. The instantaneous boundary condition on the swimmer surface defining the swimming gait is given by 
\begin{equation}\label{match}
\u^S(\x^S,t)=\frac{\p \x^S}{\p t}\cdot
\end{equation}
The Lagrangian partial derivative on the right-hand side of Eq.~\eqref{match} is order $\epsilon$ while the Eulerian velocity on the left-hand side of the equation contains terms at all order in $\epsilon$ since it is evaluated on a moving shape defined by Eq.~\eqref{xS}.  A Taylor expansion of Eq.~\eqref{match} up to order $\epsilon^2$ allows us to obtain the two boundary conditions  as
\begin{subeqnarray}
\slabel{uS1}\u_1 & = &   \U_1  
+  \O_1 \times \x^S_0  +
 \u^S_1 ,\\
\u_2 & = &  \U_2 
+  \O_2\times \x^S_0
+   \u^S_2,
\slabel{uS2}
\end{subeqnarray}
where 
$\u^S_1={\p \x_1^S}/{\p t } |_{\x^S_0}$
and
$\u^S_2=-\x_1^S\cdot\nabla \u_1|_{\x^S_0} +  \O_1  \times   \x^S_1
$. 

A final important point to note  is that since we are using an approach in domain perturbation, all fields are defined with boundary conditions on the $O(\epsilon^0)$ shape $S_0$.  This shape is fixed in time, a fact which as we see below is critical.

\subsection{Constitutive relationship}
For this integral theorem, we place no restriction on the Deborah number for the flow, and will allow the period of the surface motion to be on the same order as the fluid relaxation time, but the small value of 
 $\epsilon$ will ensure that the Weissenberg number remains small. We consider fluids obeying a general, multi-mode,  differential relationship with a spectrum of relaxation times in which  the deviatoric stress, $\btau=\bs+p \bf 1$, is written as a sum
\begin{equation}\label{const0}
\btau=\sum_{i}\btau^i.
\end{equation}
Each stress, $\btau^i$, is assumed to be following a nonlinear evolution equation of the form 
\begin{equation}\label{const}
(1+ \A_i ) \btau^i  +  \M_i(\btau^i ,\u) = \eta_i(1+\B_i)\gamd+\N_i(\gamd ,\u),
\end{equation}
where the repeated indices $i$ do not imply  Einstein summations. In Eq.~\eqref{const} $\A_i$  and $\B_i$ are arbitrary linear differential operators in time (for example a time scale times a time derivative giving Maxwell-like terms);  the symmetric tensors $\M_i$ and $\N_i$ are  arbitrary nonlinear  differential operators in space (for example, upper-convective derivatives) which are differentiable  and contain  no linear part (so at least quadratic); and $\eta_i$ is the zero-shear rate viscosity of the $i^{th}$ mode.

The assumed constitutive relationship,  
Eqs.~\eqref{const0}-\eqref{const}, is very general, and includes all classical non-Newtonian models from continuum mechanics, including all  Oldroyd-like models (upper- and lower-convected Maxwell, corotational Maxwell and  Oldroyd, Oldroyd-A and -B, Oldroyd 8-constant model, Johnson-Segalman-Oldroyd),  Giesekus and  Phan-Thien-Tanner nonlinear polymeric models, the second and $n^{th}$ order fluid approximation, all generalized Newtonian fluids, and all  multi-mode version of these constitutive models \cite{oldroyd1950,bird76,birdvol1,tanner88,bird95,larson99,morrison}. 
Furthermore,  although  the FENE-P constitutive relationship does not  exactly take the form in Eqs.~\eqref{const0}-\eqref{const}, it agrees with it for small deformations   
\cite{lauga07}, so our approach is valid for the FENE class of models  too.

\subsection{First order solution}
\label{firstosol}
At leading order, the constitutive equation for each mode is linearized and becomes 
\begin{equation}\label{order1_i}
(1+ \A_i ) \btau^i_1  = \eta_i(1+\B_i)\gamd_1.
\end{equation}
For each mode, we obtain therefore a  linearly viscoelastic fluid on a fixed shape, $S_0$, a problem which was almost already solved in \ref{sec:LVFS}.

In order to proceed in the analysis we will make the assumption, relevant to all small-scale biological swimmers, that the shape change occurs periodically in time with a fixed period, denoted $T$. We thus  use Fourier series, and we write for all  functions $h$ of period $T=2\pi/\omega$
\begin{equation}
h(t)=\sum_{n=-\infty}^\infty \tilde h ^{(n)}e^{i n \omega t}
,\quad 
\tilde h^{(n)}= \frac{1}{T}\int_0^T h (t)e^{-i n \omega t}\,\d t.
\end{equation}
Evaluating Eq.~\eqref{order1_i} in Fourier space leads to 
\begin{equation}\label{55}
[1+{\cal A}_i(n)]\tilde \btau^{i,(n)}_1 (\x)
=
\eta_i 
[1+{\cal B}_i(n)]
\tilde
\gamd_1^{(n)}
(\x),
\end{equation}
where ${\cal A}_i(n)$ and ${\cal B}_i(n)$ are  multiplicative operators obtained by evaluating the differential operators ${\cal A}_i$ and ${\cal B}_i$ in Fourier space. We can write Eq.~\eqref{55} compactly as
\begin{equation}
\tilde \btau^{i,(n)}_1 (\x)
=
{\cal G}_i(n)
\tilde
\gamd_1^{(n)}
(\x),
\end{equation}
where
\begin{equation}
{\cal G}_i(n)=\eta_i \frac{1+{\cal B}_i(n)}{1+{\cal A}_i(n)}\cdot
\end{equation}
Summing on all the modes $i$ we then obtain the Fourier components of the total stress as Newtonian-like
\begin{equation}\label{order1_const}
\tilde \btau^{(n)}_1 (\x)
=
{\cal G}(n)
\tilde
\gamd_1^{(n)}
(\x),
\end{equation}
with effective complex viscosity 
\begin{equation}\label{eff_visc}
{\cal G}(n)=\sum_i{\cal G}_i(n).
\end{equation}

To within a rescaling of the pressure, the problem posed by Eq.~\eqref{order1_const} is that of force- and torque-free swimming a linear viscoelastic fluid with a surface velocity defined on a fixed shape, $S_0$. This is therefore the same problem as in   \S\ref{sec:LVFS}, and thus the  swimming kinematics at order $\epsilon$ are the same as the Newtonian one and we obtain for each Fourier component 
\begin{equation}
\hat \F\cdot\tilde \U_1^{(n)} +\hat \L \cdot \tilde\O_1^{(n)} = -\int\!\!\!\int _{S_0}\n_0\cdot\hat \bs \cdot\tilde \u^{S,(n)}_1\,\d S
\label{order1}.
\end{equation}

Given that the shape $S_0$ does not vary with time, one  can invert the Fourier transform in Eq.~\eqref{order1} to obtain
\begin{equation}\label{order1_final}
\hat \F\cdot \U_1 +\hat \L \cdot \O_1 = -\int\!\!\!\int _{S_0}\n_0\cdot\hat \bs \cdot \u^S_1\,\d S.
\end{equation}
Notice that, similarly to the problem addressed in \S\ref{sec:LVFS}, all  material properties of the fluid have disappeared at leading order. They will however matter  at next order. 

The result of Eq.~\eqref{order1_final} can also be used to show that the time-averaged locomotion at leading order is always zero. From Eq.~\eqref{uS1} we see that $ \u_1^S$ is an exact time-derivative. We therefore have $\langle \u_1^S(\x^S_0,t)\rangle ={\bf 0}$ and thus taking the time-average of Eq.~\eqref{order1_final} leads to 
\begin{equation}\label{order1_average}
\hat \F\cdot \langle\U_1 \rangle +\hat \L \cdot \langle \O_1 \rangle = 0,
\end{equation}
and therefore $\langle \U_1\rangle=\langle \O_1 \rangle={\bf 0}.$
Similarly to the Newtonian case, net swimming 
occurs therefore at order $\epsilon^2$ at least 
\cite{brennen77,taylor51,lauga07}.

\subsection{Second-order solution}\label{secondo}
We now consider the expansion at second order.

\subsubsection{Constitutive relationship}
The constitutive relationship, Eq.~\eqref{const},  is  written at order  $\epsilon^2$ as
\begin{equation}\label{const_order2}
(1+ \A_i ) \btau^i_2   = \eta_i(1+\B_i)\gamd_2+\H_i[\u_1].
\end{equation}
{Unlike the expansion considered in \S\ref{sec:weakly} for weakly non-Newtonian flows, the general model considered in this section does allow for history terms in the evolution of the fluid stress (${\cal A}_i \neq 0$) and thus the problem  requires us to consider each Fourier mode separately.} In Eq.~\eqref{const_order2}, the nonlinear operator, $\H_i$, is only a {functional} of $\u_1$ and is formally written using gradients in the operators $\N_i$ and $\M_i$  as
\begin{equation}
\H_i[\u_1]=
\gamd_1:
\left[(\nabla_{\gamd} \nabla_\u 
\N_i)\big\vert_{{\bf 0},{\bf 0}}\right]\cdot \u_1
-\btau_1^i: 
\left[(\nabla_{\btau^i} \nabla_\u 
\M_i)\big\vert_{{\bf 0},{\bf 0}}\right]\cdot \u_1
,
\quad
\end{equation}
with the relationship between $\btau^i_1$ and $\gamd_1$ given by Eq.~\eqref{order1_i}, and where we recall that $ \gamd_1=\nabla \u_1 + \nabla \u_1^T$. Using Fourier series and using the same notation as in \S\ref{firstosol}, we can then rewrite  Eq.~\eqref{const_order2} as 
\begin{equation}
[1+{\cal A}_i(n)] \tilde \btau_2^{i,(n)}
(\x)=\eta_i[1+{\cal B}_i(n)]\tilde \gamd_2^{(n)}(\x) + \widetilde{\H_i[\u_1]}^{(n)}(\x)
,
\end{equation}
or
\begin{equation}\label{order2_fourier}
 \tilde \btau_2^{i,(n)}(\x)={\cal G}_i(n)\tilde \gamd_2^{(n)}(\x) + \frac{1}{[1+{\cal A}_i(n)]}\widetilde{\H_i[\u_1]}^{(n)}(\x)
.
\end{equation}
Summing up Eq.~\eqref{order2_fourier} for all indices $i$, we obtain explicitly the second order deviatoric stress as 
\begin{equation}\label{order2_fourier_final}
 \tilde \btau^{(n)}_2(\x)={\cal G}(n)\tilde \gamd_2^{(n)}(\x) + \widetilde{ \BS [\u_1]}^{(n)}(\x),
 \end{equation}
 where we have defined
\begin{equation}
\widetilde{ \BS [\u_1]}^{(n)}(\x)
=\sum_i \frac{1}{1+{\cal A}_i(n)}\widetilde{\H_i[\u_1]}^{(n)}(\x)
\end{equation}
\subsubsection{Principle of virtual work}
After Eq.~\eqref{order2_fourier_final} we see that total stress in the fluid is given by
\begin{equation}\label{stressorder2}
\tilde \bs_2 ^{(n)}(\x) = - \tilde p_2^{(n)} (\x)  {\bf 1}+ {\cal G}(n)  \tilde \gamd_2^{(n)}(\x) +\widetilde{ \BS [\u_1]}^{(n)}(\x). 
\end{equation}
We now apply the principle of virtual work to the 
$\{\tilde \u_2^{(n)},\tilde \bs_2 ^{(n)}\}$ problem, together with a  solid body motion which takes place with the viscosity ${\cal G}(n)$, which we denote 
$\{\hat\u_{(n)},\hat\bs_{(n)}\}$\footnote{The flow field $\hat\u_{(n)}$ is  not a Fourier component nor a series expansion: the subscript  ${(n)}$ is used as a reminder that the associated viscosity is  ${\cal G}(n)$.}. The solid body motion  is associated with  complex forces and torques given by 
$\hat \F_{(n)}$ and $\hat \L_{(n)}$, resulting in solid body kinematics given by  $\hat\U_{(n)}$ and $\hat\O_{(n)}$. As a difference with the calculation in \S\ref{sec:LVFS}, here the value of the complex viscosity matters and thus the solid body motion in the hat problem will always be a function of the 
order, $n$, of the Fourier mode  considered (hence the notation chosen).

Since both problems satisfy that the divergence of the stress tensor is zero, we compute the virtual work and obtain 
\begin{equation}
\hat \u_{(n)} \cdot \nabla\cdot \tilde \bs_2^{(n)} = \tilde \u_2^{(n)} \cdot \nabla \cdot \hat\bs_{(n)},
\end{equation}
which we then integrate in the entire fluid  volume  and  use the divergence theorem to obtain
\begin{equation}\label{neworder2}
\int\!\!\!\int _{S_0}\n_0\cdot \hat\bs_{(n)}\cdot \tilde \u_2^{(n)}\,\d S
-
\int\!\!\!\int _{S_0}\n_0\cdot \tilde  \bs_2^{(n)} \cdot \hat\u_{(n)}\,\d S
=
\int\!\!\!\int\!\!\!\int_{V_0} \tilde \bs_2^{(n)}:\nabla \hat\u_{(n)} \, \d V
-
\int\!\!\!\int\!\!\!\int_{V_0} \hat\bs_{(n)}:\nabla \tilde \u_2 ^{(n)} \, \d V
.\end{equation}
We then plug Eq.~\eqref{stressorder2} into the right-hand side of Eq.~\eqref{neworder2} to get
\begin{equation}
\int\!\!\!\int\!\!\!\int_{V_0} \tilde \bs_2^{(n)}:\nabla \hat\u_{(n)} \, \d V
-
\int\!\!\!\int\!\!\!\int_{V_0} \hat\bs_{(n)}:\nabla \tilde \u_2^{(n)}  \, \d V
=\int\!\!\!\int\!\!\!\int_{V_0}\widetilde {\BS[\u_1]} ^{(n)}: \nabla \hat\u_{(n)} \, \d V,
\end{equation}
where the symmetric terms have disappeared due to incompressibility and by equality of their  viscosity, similarly to Eq.~\eqref{goestozero}, 
 so that we obtain
\begin{equation}\label{middle}
\int\!\!\!\int _{S_0}\n_0\cdot \hat\bs_{(n)}\cdot \tilde \u_2^{(n)}\,\d S
-
\int\!\!\!\int _{S_0}\n_0\cdot \tilde  \bs_2^{(n)}\cdot \hat\u_{(n)}\,\d S
=
\int\!\!\!\int\!\!\!\int_{V_0}\widetilde {\BS[\u_1]} ^{(n)}: \nabla \hat\u_{(n)} \, \d V
.\end{equation}
On the first left-hand side of Eq.~\eqref{neworder2} we write, on  $S_0$, the Fourier components of the boundary condition at order $\epsilon^2$, namely 
$\tilde \u_2^{(n)}(\x_0^S) =\tilde \U_2^{(n)}+ \tilde \O_2^{(n)}\times \x_0^S + \tilde  \u_2^{S,(n)}(\x_0^S)$, so that the integral formulation, Eq.~\eqref{middle}, becomes
\begin{eqnarray}\notag
\hat \F_{(n)}\cdot \tilde  \U_2^{(n)}  +\hat \L_{(n)} \cdot \tilde   \O_2^{(n)} &=&
- \int\!\!\!\int _{S_0}\n_0\cdot\hat \bs_{(n)} \cdot \tilde \u_2^{S,(n)} \,\d S
+\int\!\!\!\int _{S_0}\n_0\cdot \tilde  \bs_2^{(n)}\cdot \hat\u_{(n)}\,\d S
\\
&&+\int\!\!\!\int\!\!\!\int_{V_0} \widetilde {\BS[\u_1]}^{(n)} : \nabla \hat\u _{(n)}\, \d V,
\label{almostthere}
\end{eqnarray}
where $\hat \F_{(n)}$ and $\hat \L_{(n)}$ depend on $n$ through the complex viscosity $\G(n)$. The final term we have to evaluate in Eq.~\eqref{almostthere} is the integral
\begin{equation}\label{eqI}
\I=\int\!\!\!\int _{S_0}\n_0\cdot \tilde  \bs_2^{(n)}\cdot \hat\u_{(n)}\,\d S,
\end{equation}
and since the boundary condition for the hat problem   on the surface is $\hat \u_{(n)} = \hat\U_{(n)} + \hat\O_{(n)}\times \x_0^S $, $\I$ is given by
\begin{equation}\label{92}
\I=\left[\int\!\!\!\int _{S_0}\n_0\cdot \tilde  \bs_2^{(n)}\,\d S\right]\cdot \hat\U_{(n)}
+
\left[\int\!\!\!\int _{S_0}\x_0^S\times(\n_0\cdot \tilde  \bs_2^{(n)})\,\d S\right]\cdot \hat\O_{(n)}.
\end{equation}
The terms multiplying the solid-body kinematics in  Eq.~\eqref{92} seem to involve the  $O(\epsilon^2)$ forces and torques on the swimmer. In the next section we show how to use arguments of vector calculus and differential geometry to evaluate them explicitly.  

\subsubsection{Differential geometry} \label{diffgeom}

Since we are using a domain expansion method, particular attention needs to be paid to the expressions for the   hydrodynamic forces and moments acting  on the swimmer. Indeed,  these are to be evaluated on a shape changing in time, and thus the application of the force- and moment-free condition is not  straightforward.

Since motion of the swimmer tangential to its surface does not lead to changes in its shape, only the normal component of the surface motion will contribute.  We thus write the shape variation of the  periodically moving interface, $S(t)$, as the normal projection to the motion of the material points, and thus we describe the surface as   
$\x=\x_0^S + \epsilon \delta_1 (\x_0^S,t) \n_0(\x_0^S) $, where $\n_0$ is the normal to the surface $S_0$ at point $\x_0^S$, and the function $\delta_1$, with units of length, represents the normal shape deformation of the reference surface. Given that we have material points whose dynamics is given by Eq.~\eqref{xS} we necessarily have $\delta_1=
\x_1^S\cdot \n_0$.  Note that for a squirming motion, we have by definition $\delta_1=0$, so $\x=\x_0^S$ and thus $S(t)=S_0$ for all times. Associated with this shape variation is the normal to the surface, which is expanded as $\n=\n_0(\x_0^S) + \epsilon \n_1(\x_0^S)+...$, with all fields described on the undeformed surface, $S_0$. On the swimmer surface we thus have the expansion
\begin{equation}\label{bs_exp}
\n\cdot \bs = \epsilon \n_0\cdot \bs_1 + \epsilon^2(\n_0\cdot \bs_2 + \n_1\cdot \bs_1).
\end{equation}

Using this description, we can calculate the asymptotic value of the surface integral $W$ of an arbitrary scalar field $w(\x)$
\begin{equation}
W=\int\!\!\!\int _{S(t)} w(\x)\,\d S.
\end{equation}
Expanding the integrand as $w(\x)=\epsilon w_1(\x)+\epsilon^2 w_2(\x) + ...$ and using Taylor expansion to evaluate the integral on the reference $S_0$ we obtain
$W=\epsilon W_1 + \epsilon^2 W_2 +...$
with
\begin{equation}\label{53}
W_1=\int\!\!\!\int _{S_0}  w_1(\x_0^S)\,\d S,\quad {\rm and}\quad
W_2=\int\!\!\!\int _{S_0}  \left(w_2 +  \delta_1\frac{\p w_1}{\p n}  \right) (\x_0^S)\,\d S
\end{equation}
where the normal derivative is understood as normal to the unperturbed surface, ${\it i.e.}$,  $\p w_1/{\p n}=\n_0\cdot \nabla w_1$. 

The force and torque on the swimmer are formally given by the integrals
\begin{equation}
\F=\int\!\!\!\int _{S(t)}\n\cdot\bs \,\d S,\quad
\O=\int\!\!\!\int _{S(t)}\x \times (\n\cdot\bs) \,\d S,
\end{equation}
for which we will have the expansion
\begin{equation}
\{\F,\L\}=\epsilon\{\F_1,\L_1\}+\epsilon^2\{\F_2,\O_2\}+...
\end{equation}
with the forces and torques equal to zero at each order. Applying the results above with $w$ equal to each to each component of the force per unit area on the swimmer, $\bs\cdot \n$, expanded as in Eq.~\eqref{bs_exp} we obtain at fist order the expected integrals
\begin{subeqnarray}\label{FO_order1}
\F_1  &=&\int\!\!\!\int _{S_0}\n_0\cdot \bs_1 \,\d S={\bf 0},\quad\\
 \L_1 &=&\int\!\!\!\int _{S_0}\x_0^S \times (\n_0\cdot \bs_1 ) \,\d S={\bf 0}
\end{subeqnarray}
while at order $\epsilon^2$ it leads to additional terms and
\begin{subeqnarray}\label{diff1}
\F_2 & = & 
\slabel{diff1F}\int\!\!\!\int _{S_0}
\left( \n_0\cdot \bs_2 + \n_1\cdot \bs_1 + 
\delta_1 
\n_0\cdot \frac{\partial \bs_1}{\partial n}  
\right)\,\d S = {\bf 0},
\\
\slabel{diff1L}\L_2 & = & 
\int\!\!\!\int _{S_0}
\x_0^S\times \left( \n_0\cdot \bs_2 + \n_1\cdot \bs_1 + 
\delta_1 
\n_0\cdot \frac{\partial \bs_1}{\partial n}  
\right)\,\d S
+\int\!\!\!\int _{S_0}\delta_1  \n_0 \times (\n_0\cdot \bs_1 )
\,\d S
=  {\bf 0},\quad\quad 
\end{subeqnarray}
for all times.

We can now use   differential geometry and vector calculus to simplify the results in Eq.~\eqref{diff1}. Given that the surface shape is described by   $\x=\x_0^S + \epsilon \delta_1 (\x_0^S,t) \n_0(\x_0^S) $ then it is straightforward to see that the first perturbation of the surface normal, $\n_1$, is given by minus the surface gradient of the shape field $\delta_1$, i.e.~$\n_1(\x_0^S)=-\nabla_{\x_0^S} \delta_1$. In Eq.~\eqref{diff1} we therefore have
\begin{equation}\label{58}
\n_1\cdot \bs_1 + 
\delta_1 
\n_0\cdot \frac{\partial \bs_1}{\partial n}  
=-(\nabla_{\x_0^S} \delta_1)\cdot\bs_1 + \delta_1 
\n_0\cdot \frac{\partial \bs_1}{\partial n}  \cdot
\end{equation}
We can then use the identity from vector calculus
\begin{equation}
\nabla_{\x_0^S} \cdot (\delta_1\bs_1) =
\delta_1(\nabla_{\x_0^S} \cdot \bs_1 )+ 
(\nabla_{\x_0^S} \delta_1)\cdot\bs_1
\end{equation}
to simplify Eq.~\eqref{58} into
\begin{equation}\label{60}
\n_1\cdot \bs_1 + 
\delta_1 
\n_0\cdot \frac{\partial \bs_1}{\partial n}  
=
- \nabla_{\x_0^S} (\delta_1\cdot\bs_1) 
+\delta_1\left(\nabla_{\x_0^S} \cdot \bs_1  + 
\n_0\cdot \frac{\partial \bs_1}{\partial n} \right) \cdot
\end{equation}
The last term in parenthesis in Eq.~\eqref{60} is an expression for the three-dimensional divergence of $\bs_1$, which is zero, 
\begin{equation}
\nabla_{\x_0^S} \cdot \bs_1  + 
\n_0\cdot \frac{\partial \bs_1}{\partial n}=\nabla\cdot \bs_1=\bf 0
\end{equation}
since the flow at each order in the perturbation expansion satisfy Cauchy's equation of motion, $\nabla\cdot \bs_i=\bf 0$. This results allows us to simplify each expression in  Eq.~\eqref{diff1}. Starting with the force in Eq.~\eqref{diff1F}, we now have
\begin{equation}\label{diff2F}
\F_2 = 
\int\!\!\!\int _{S_0}
\left[ \n_0\cdot \bs_2 
- \nabla_{\x_0^S} \cdot (\delta_1\bs_1) \right]\,\d S = {\bf 0},
\end{equation}
The integral of the second term in Eq.~\eqref{diff2F} is a surface divergence integrated on a closed surface, and therefore equal to zero (this can be viewed as an application of the curl theorem). And therefore we finally obtain the simple expression for the second-order force as
\begin{equation}\label{diff3F}
\F_2 = 
\int\!\!\!\int _{S_0}
 \n_0\cdot \bs_2 
\,\d S = {\bf 0}.
\end{equation}

The equation for the moment, Eq.~\eqref{diff1L}, is now written as
\begin{equation}\label{diff2L}
\L_2 = 
\int\!\!\!\int _{S_0}
\x_0^S\times ( \n_0\cdot \bs_2 )
\,\d S
+\int\!\!\!\int _{S_0}
\left[
\delta_1  \n_0 \times (\n_0\cdot \bs_1 )
-\x_0^S\times  \nabla_{\x_0} \cdot (\delta_1\bs_1) 
\,\d S\right]
=  {\bf 0}. 
\end{equation}
Let us now show that  the second integral in 
Eq.~\eqref{diff2L} is identically zero. If the shape of the swimmer does not vary, then $\delta_1=0$ and that second integral is trivially equal to zero. If the shape of the swimmer does change in time, then since we have freedom in how we define the reference shape $S_0$,  we can always change how we parametrize it thus without loss of generality can assume  $S_0$ is locally flat. We then employ cartesian coordinates  with $\n_0= \e_z$ and the surface defined as $z= 0$, so that $\x_0^S=x\e_x+y\e_y$. In that case, the first integrand  in the  second integral in Eq.~\eqref{diff2L} is given by 
\begin{equation}\label{91}
\delta_1 \n_0\times(\n_0\cdot \bs_1) 
=\delta_1\e_z\times(\sigma_{1,xz} \e_x +\sigma_{1,yz} \e_y )
=\delta_1(\sigma_{1,xz} \e_y -\sigma_{1,yz} \e_x ).
\end{equation}
The surface divergence in second integrand is given by 
\begin{equation}\label{second}
  \nabla_{\x_0^S} \cdot (\delta_1\bs_1) 
= \e_\alpha\p_\alpha \cdot (\delta_1\sigma_{1,ij}\e_i\e_j) 
= \p_\alpha  (\delta_1\sigma_{1,\alpha j})\,\e_j
\end{equation}
where we have used the convention that Einstein's summation notation with Latin letters ($i$, $j$,...) implies a summation on all three coordinates $x$, $y$, $z$ while a summation with Greek letters ($\alpha$, $\beta$, ...) implies a summation only on the surface coordinates  $x$ and $y$. Using Eq.~\eqref{second} we can then compute explicitly the second integrand as
\begin{eqnarray}
-\x_0^S\times  \nabla_{\x_0^S} \cdot (\delta_1\bs_1) 
= -x_\beta \e_\beta\times  \p_\alpha  (\delta_1\sigma_{1,\alpha j})\e_j
= -\epsilon_{m\beta j}x_\beta   \p_\alpha  (\delta_1\sigma_{1,\alpha j})\, \e_m.
\end{eqnarray}
In order to take force that term to take the form of a surface divergence, we can re-write it as
\begin{equation}\label{94}
-\x_0^S\times  \nabla_{\x_0^S} \cdot (\delta_1\bs_1) 
= -\p_\alpha  (\epsilon_{m\beta j}x_\beta   \delta_1\sigma_{1,\alpha j})\, \e_m+
\epsilon_{m\alpha j}   \delta_1\sigma_{1,\alpha j}\, \e_m.
\end{equation}
The first term on the right-hand side of Eq.~\eqref{94} is a surface divergence and will thus disappear when integrate on the close surface $S_0$. The second term can be evaluated explicitly because for all indices $j$ equal to $x$ or $y$, since the tensor $\bs_1$ is symmetric and the tensor $\boldsymbol \epsilon$ is antisymmetric, terms with $(\alpha, j)$ and $(j,\alpha)$ will cancel out, and thus only the terms with $j=z$ survive.  This leads to 
\begin{equation}\label{95}
\epsilon_{m\alpha j}   \delta_1\sigma_{1,\alpha j}\, \e_m=
\epsilon_{m\alpha z}   \delta_1\sigma_{1,\alpha z}\, \e_m=
   \delta_1(\sigma_{1,y z}\, \e_x
-   \sigma_{1,x z}\, \e_y).
\end{equation}
We then see that the result of Eq.~\eqref{95} exactly cancels out the first integrand given in Eq.~\eqref{91} and therefore the whole second integral in Eq.~\eqref{diff2L} disappears, leaving the second-order moment to be given by
\begin{equation}\label{diff2L_2}
\L_2 = 
\int\!\!\!\int _{S_0}
\x_0^S\times ( \n_0\cdot \bs_2 )
\,\d S.
\end{equation}

\subsubsection{Integral theorem}

Using the results from the previous section and enforcing that swimming is force- and torque-free at order two, $\F_2=\L_2={\bf 0}$, we obtain simply
\begin{equation}
\int\!\!\!\int _{S_0}\n_0\cdot \bs_2 \,\d S = 
{\bf 0}
, \quad 
\int\!\!\!\int _{S_0}\x_0^S\times( \n_0\cdot \bs_2)\,\d S  =  {\bf 0}
.
\end{equation}
In Fourier space, since the reference shape $S_0$ is fixed, we obtain for each Fourier component 
\begin{equation}
\int\!\!\!\int _{S_0} \n_0\cdot \tilde  \bs_2^{(n)}  \,\d S  =  {\bf 0},\quad 
\int\!\!\!\int _{S_0}\x_0^S\times( \n_0\cdot \tilde  \bs_2^{(n)}) \,\d S   =  
{\bf 0}.
\end{equation}
From Eq.~\eqref{eqI}, we then obtain $\I=0$, and 
 Eq.~\eqref{almostthere} leads then to the final integral theorem 
\begin{equation}
\hat \F_{(n)}\cdot \tilde  \U_2^{(n)}  +\hat \L_{(n)} \cdot \tilde   \O_2^{(n)} =
- \int\!\!\!\int _{S_0}\n_0\cdot\hat \bs_{(n)} \cdot \tilde \u_2^{S,(n)} \,\d S
+\int\!\!\!\int\!\!\!\int_{V_0} \widetilde {\BS[\u_1]}^{(n)} : 
\nabla \hat\u _{(n)}\, \d V
\label{integral_3}.
\end{equation}

Our final result, Eq.~\eqref{integral_3}, provides explicit expressions for the Fourier modes of the swimming kinematics at order 2, namely $\U_2^{(n)}$ and $\O_2^{(n)}$,  allowing to 
 reconstruct the whole time-dependent   swimming velocity, $\U_2$, and rotation rate, $\O_2$, at order $O(\epsilon^2)$. {This is the most important result from our paper}.  Physically, we see that the swimming kinematics are simply given by the sum of a Newtonian component and a non-Newtonian part. Since the constitutive relationship has been left very general, the result in Eq.~\eqref{integral_3} is expected to be applicable to a wide range of complex fluids, {swimmer geometry and deformation kinematics}.
 
In order to mathematically evaluate Eq.~\eqref{integral_3}, we see that the following knowledge is required. We see to know the full  velocity field at order 1, $\u_1$, the Fourier component of the second-order swimming gait, $\tilde \u_2^{S,(n)}$,  and a dual Newtonian solution, $\{\hat\u_{(n)},\hat\bs_{(n)}\}$, corresponding to solid body motion with net force $\F_{(n)}$ and moment $\O_{(n)}$. 
The dual Newtonian problem corresponds to rigid-body motion in a Newtonian fluid of complex viscosity  ${\cal G}(n)$, and can be deduced, by exploiting the linearity of Stokes equations, from the flow at a reference viscosity by a simple rescaling. The order 1 swimming problem, $\u_1$, has known boundary conditions computed in Eq.~\eqref{order1_final}, and has therefore the computational complexity of a Newtonian problem. Similarly to the previous theorem, the gradient $\nabla \hat\u _{(n)}$ 
in Eq.~\eqref{integral_3} can be replaced by the symmetric part of the velocity gradient, giving the alternative form
 \begin{equation}
\hat \F_{(n)}\cdot \tilde  \U_2^{(n)}  +\hat \L_{(n)} \cdot \tilde   \O_2^{(n)} =
- \int\!\!\!\int _{S_0}\n_0\cdot\hat \bs_{(n)} \cdot \tilde \u_2^{S,(n)} \,\d S
+\int\!\!\!\int\!\!\!\int_{V_0} \widetilde {\BS[\u_1]}^{(n)} : \nabla \hat\e _{(n)}\, \d V
\label{integral_3_bis}.
\end{equation}

\subsection{Time-averaged swimming kinematics}

The most important component of the swimming kinematics is the  $n=0$ Fourier mode giving access to the time-average of the motion.    In that case, the dual Newtonian problem in Eq.~\eqref{order1_final}, $\hat\u$,  occurs with viscosity $\G(n=0)=\sum_i \eta_i \equiv \eta $.  Using the notation $\langle  f \rangle=\tilde{f}^{(0)}$, to denote time averaging, the integral formula giving the time-averaged swimming kinematics is given by
\begin{equation}\label{timeav}
\hat \F\cdot\langle \U_2 \rangle+\hat \L \cdot \langle\O_2\rangle =
- \int\!\!\!\int _{S_0}\n_0\cdot\hat \bs \cdot \langle \u_2^S\rangle \,\d S
+\int\!\!\!\int\!\!\!\int_{V_0}\langle \BS[\u_1]\rangle :  \hat\e \, \d V
\end{equation}

\subsection{Locomotion of a sphere}

A special case of interest for exact calculations is that of a swimming of a spherical body of radius $a$. This is the Lighthill and Blake model \cite{Lighthill1952,Blake1971a} addressed in  \S\ref{sec:LVFS}.

Inside the fluid, we have the velocity field given by 
\begin{equation}\label{103}
\hat\u= \frac{3}{4}a\left[\frac{\bf 1}{r}+\frac{\r\r}{r^3}\right] \cdot \hat \U+ \frac{1}{4}a^3 \left[\frac{\bf 1}{r^3}-\frac{3\r\r}{r^5}\right]\cdot \hat \U
 +\frac{a^3}{r^3}\hat\O\times \r,
\end{equation}
with boundary conditions $\hat\u=\hat\U + \O\times\x_0^S$ on the sphere. The surface stress then takes the form 
\begin{equation}
\n_0\cdot \hat \bs=-\frac{3\eta}{2a}\hat \U - {3\eta} \hat\O\times \n_0.
\end{equation}
 In that case, and focusing on the time-averaged locomotion, Eq.~\eqref{timeav} becomes 
\begin{equation}\label{104}
\hat \F\cdot\langle \U_2 \rangle+\hat \L \cdot \langle\O_2\rangle =
6\pi a \eta \, \hat \U\cdot
\overline{ \langle \u_2^S\rangle}
+ 8\pi a^3 \eta \, \hat \O\cdot
(\overline{ \x_0^S\times \langle \u_2^S\rangle})
+\int\!\!\!\int\!\!\!\int_{V_0}\langle \BS[\u_1]\rangle :  \hat\e \, \d V,
\end{equation}
where overline indicates surface average 
$\overline{ w}=( \int\!\!\!\int _{S_0}w \,\d S)/({4\pi a^2})$. We have  $\hat \F=-6\pi \eta a \hat \U$ and  $\hat \L=-8\pi \eta a^3 \hat \O$. The hat flow field in Eq.~\eqref{103} can be formally written as 
$\hat\u=\hat\P\cdot \hat\U + \hat\Q\cdot\hat\O $
leading  to $\hat\e= \hat\E(\hat\P).\hat\U+ \hat\E(\hat\Q).\hat\O$ using the definition for, an arbitrary second-order tensor, $\T$, of the third order tensor
$ \{\hat\E(\hat\T)\}_{ijk}=\frac{1}{2}(\p_i\hat T_{jk}+\p_j\hat T_{ik})$. 
Considering separately $\hat \U=\bf 0$ and $\hat \O=\bf 0$ 
we then  obtain
from Eq.~\eqref{104}
\begin{eqnarray}
\langle \U_2 \rangle\label{U2_sphere}
 & = & -
\overline{ \langle \u_2^S\rangle}
-\frac{1}{6\pi\eta a}\int\!\!\!\int\!\!\!\int_{V_0}\langle \BS[\u_1]\rangle :  \hat\E(\hat \P) \, \d V,\\
\langle\O_2\rangle &=& - \overline{ \x_0^S\times \langle \u_2^S\rangle}
-\frac{1}{8\pi\eta a^3}\int\!\!\!\int\!\!\!\int_{V_0}\langle \BS[\u_1]\rangle :  \hat\E (\hat\Q) \, \d V,
\end{eqnarray}
with similar formulae available for each of the Fourier modes (modulo the correct definition of the complex viscosity for mode $n$).

\section{Application to the scallop theorem}
\label{sec:scallop}
In addition to allowing the calculation of non-Newtonian swimming of biological and  synthetic swimmers, our integral theorem allows us to formally revisit Purcell's scallop theorem \cite{purcell1977} in the context of complex fluids. That theorem states that deformations which are not identical under a time-reversal symmetry (so-called non-reciprocal) are required to induce locomotion in  Newtonian Stokes flows. Using the formalism of the Newtonian integral theorems from \S\ref{sec:Newtonian}, Eq.~\eqref{N_final}, reciprocal deformations are those for which $\langle \u^S\rangle={\bf 0}$ leading to $\langle \U\rangle=\langle \O\rangle=\bf 0$.

When considering the scallop theorem in non-Newtonian flows, two distinct points need to be addressed. The first is answering the question: Is the scallop theorem still valid in general? The answer is obviously no. Fluids with nonlinear rheology can be exploited to generate propulsion from time-reversible actuation  \cite{normand08,pak10,pak12,keim12}. The simplest way to see this from our results is to realize that the operators $\BS[\u]$ appearing in \S\ref{sec:small} (Eq.~\ref{integral_2}) and \S\ref{sec:weakly} (Eq.~\ref{integral_3}) are nonlinear operators acting on the flow field at the previous order. If that flow includes a time-varying component $\propto e^{i\omega t}$ induced by the time-reversible motion, then  $\BS[\u]$ will generate harmonics, with in general a nonzero time-average. A specific example will be given in the next section.

A second, more interesting point, is whether there exists a categories of non-Newtonian fluids for which the scallop theorem would be remain valid. Our integral theorems can be used to show that  for any linearly viscoelastic fluid a time-reversible actuation cannot lead to any net motion. In the case where the surface actuation  is tangential to the swimmer surface, as addressed in \S\ref{sec:LVFS}, we obtain by simply applying Eq.~\eqref{integral_1} in the reciprocal case that  $\hat \F\cdot \langle \U\rangle +\hat \L \cdot \langle \O\rangle =0$ and thus $\langle \U \rangle=\langle \O \rangle=\bf 0$. That result is true for arbitrary amplitude of the motion. When the surface motion includes a nonzero component normal to the shape, and thus leads to shape changes, we can apply the small-amplitude results of \S\ref{sec:small} and Eq.~\eqref{integral_3}. If the fluid is linearly viscoelastic, then we have $\BS=\bf 0$, leading to 
$\hat \F\cdot\langle \U_2 \rangle+\hat \L \cdot \langle\O_2\rangle =
0$ and therefore $\langle \U_2 \rangle=\langle \O_2 \rangle=\bf 0$. Here again we see that reciprocal swimming is not possible in a linearly viscoelastic fluid.

\section{Locomotion in an Oldroyd-B fluid}\label{sec:Old}
A model of particular interest  for the dynamics of polymeric fluids is the Oldroyd-B fluid, which can be derived formally from a dilute solution of elastic dumbbells \cite{oldroyd1950,bird76,birdvol1,tanner88,bird95,larson99,morrison}.  We show here how to apply Eq.~\eqref{integral_3} for the Oldroyd-B fluid and consider the special case of squirming motion.

\subsection{General framework} 
{
The constitutive equation for the Oldroyd-B fluid is written as 
\begin{equation}\label{OB_0}
\btau +\lambda \dbtau= (\eta_s+\eta_p) \gamd + \eta_s \lambda \dgamd ,
\end{equation}
where $\lambda$ is the relaxation time for the fluid, $\eta_s$ the solvent viscosity, and $\eta_p$ the polymeric contribution to the viscosity. 
In Eq.~\eqref{OB_0}, $\stackrel{\triangledown}{{\bf a}}$ denotes the upper convected derivative for a tensor $\bf a$
\begin{equation}
\stackrel{\triangledown}{{\bf a}}=\frac{\p {\bf a}}{\p t} + {\bf u}\cdot \nabla{\bf a} - (^t\nabla \u\cdot {\bf a} + {\bf a}\cdot \nabla \u).
\end{equation}
  Writing $\eta\equiv \eta_s + \eta_p$ for  the total viscosity of the fluid and using the notation  $\lambda_1\equiv \lambda$ and $\lambda_2\equiv \lambda\eta_s/\eta$, the constitutive law can be re-written as
\begin{equation}\label{OB}
\btau +\lambda_1 \dbtau= \eta \left(\gamd + \lambda_2 \dgamd \right),
\end{equation}
and $\lambda_2$ is referred to as the retardation time scale for the fluid. 
Note that in this model we always have $\lambda_2/\lambda_1 < 1$.
}

The expansion at order one of Eq.~\eqref{OB} leads to
\begin{equation}\label{Fourier1O}
 \btau_{1}+\lambda_1\frac{\partial\btau_{1}}{\partial t}=\gamd_{1}
 + \lambda_2\frac{\partial\gamd_{1}}{\partial t},
\end{equation}
while the second order term gives 
\begin{eqnarray}\label{order2O}
\left(1+\frac{\partial}{\partial t}\lambda_1\right)\btau_{2}
-\eta \left(1+\frac{\partial}{\partial t}\lambda_2\right)\gamd_{2} & = 
& \eta \lambda_2\left[\mathbf{u}_{1}\cdot\nabla\gamd_{1}-\left(^{t}\nabla\mathbf{u}_{1}\cdot\gamd_{1}+\gamd_{1}\cdot\nabla\mathbf{u}_{1}\right)\right]\nonumber\\
 & - & \lambda_1\left[\mathbf{u}_{1}\cdot\nabla\btau_{1}-\left(^{t}\nabla\mathbf{u}_{1}\cdot\btau_{1}+\btau_{1}\cdot\nabla\mathbf{u}_{1}\right)\right],  \end{eqnarray}
 from which all Fourier terms can be computed. If we assume to  have only one Fourier mode, $\propto e^{i\omega t}$, in the solution at order one,  then we obtain from time-averaging Eq.~\eqref{order2O} and exploiting Eq.~\eqref{Fourier1O} written in Fourier space the explicit expression for the time-averaged stress as second order as
\begin{equation}\label{finalO}
\langle\BS[\u_1]\rangle= 2 \eta(\lambda_2-\lambda_1)
{\cal R} 
\left\{\frac{1}{1+i\lambda_{1}\omega}\left[\tilde\u_1^{(1),*}\cdot\nabla\tilde\gamd_1^{(1)}-\left(^{t}\nabla\tilde\u_1^{(1),*}\cdot\tilde\gamd_1^{(1)}+\tilde\gamd_1^{(1)}\cdot\nabla\tilde\u_1^{(1),*}\right)\right]\right\},
\end{equation}
where stars denote complex conjugates and $\cal R$ the real part of a complex expression.

\subsection{Squirming motion of a sphere}

We now consider that the swimmer is a sphere undergoing  tangential squirming motion. We further assume that all surface motion is axisymmetric so that the sphere does not rotate and only swims along a straight line, with direction $\e_z$. Using cylindrical coordinates with $\theta$ the polar angle, we thus assume that its surface deforms in time as  
\begin{equation}\label{squirm_phi}
\theta=\theta_0 + \epsilon [f(\theta_0)\sin\omega t +g(\theta_0)\sin(\omega t+\phi) ].
\end{equation}
The presence of a phase $\phi$ in Eq.~\eqref{squirm_phi} allows us to combine the periodic motion of two surface modes, characterized by the functions $f$ and $g$, and includes in particular  standing and traveling waves as special cases. 

From Eq.~\eqref{squirm_phi} we can compute the surface velocity as
\begin{equation}\label{u1s_real}
\u_1^S=a \frac{\p \theta}{\p t} \e_\theta=  
a \omega [f(\theta_0)\cos\omega t
+g(\theta_0)\cos(\omega t+\phi)]\e_\theta,
\end{equation} 
with a surface gradient given by
\begin{equation}
\frac{\p \u_1^S}{\p \theta}=  
a \omega [f'(\theta_0)\cos\omega t
+g'(\theta_0)\cos(\omega t+\phi)]\e_\theta.
\end{equation}
We can then use these results to compute the surface velocity as second-order using Eq.~\eqref{uS2}   and we obtain 
\begin{eqnarray}\label{u2s_real}
\langle \u_2^S\rangle&=& - \left\langle \theta  \frac{\p \u_1^S}{\p \theta}\right\rangle 
   =\frac{a\omega}{2} \sin\phi [f(\theta_0)g'(\theta_0)-f'(\theta_0)g(\theta_0)]\e_\theta.
\end{eqnarray}

In order to  take advantage of of Blake's mathematical framework \cite{Blake1971a} we then choose the dimensionless functions
\begin{equation}
f(\theta)=\alpha  \sin\theta\cos\theta
,\quad
g(\theta)= \beta \sin \theta.
\end{equation}
From Eq.~\eqref{u2s_real} we then obtain
\begin{equation}
\langle \u_2^S\rangle = \frac{\alpha\beta}{2}a\omega \sin\phi \sin^3\theta\, \e_\theta,
\end{equation}
giving rise to average  Newtonian swimming with order-2 speed, $\langle \U_2 \rangle_{N}$, as
 \begin{equation}
\langle \U_2 \rangle_{N}=-\overline{\langle  \u_2^S\rangle} =\frac{4\alpha \beta}{15}a \omega \sin\phi\,\e_z.
\end{equation}

In order to compute the non-Newtonian correction to the swimming speed we need to compute $\u_1$ everywhere from the knowledge of $\u_1^S$. From Eq.~\eqref{finalO} we see that all we need is the Fourier component, $  \tilde\u_1$, of $\u_1$, which we obtain from Eq.~\eqref{u1s_real} as
\begin{eqnarray}
\u_1^{S}(a,\theta,t) =  
a \omega [f(\theta_0)\cos\omega t
+g(\theta_0)\cos(\omega t+\phi)]\e_\theta
=\tilde \u_1^{S,(1)} e^{i \omega t}+\tilde \u_1^{S,(-1)} e^{-i \omega t}
,\end{eqnarray} 
with
\begin{equation}
\tilde \u_1^{S,(1)}(a,\theta)= \frac{a\omega}{2} (  \alpha  \sin\theta\cos\theta + \beta e^{i\phi} \sin \theta )
\e_\theta,
\end{equation}
and $\tilde \u_1^{S,(-1)}=\tilde \u_1^{S,(1)*}$. 
This surface velocity   leads to  swimming at order one as
\begin{equation}\label{U1s}
\tilde \U_1^{(1)}=\frac{a\omega}{3}e^{i\phi} \beta \e_z.
\end{equation}
The total  velocity at the surface of the spherical swimmer in the laboratory frame, including the component from swimming, Eq.~\eqref{U1s}, is thus given by
\begin{eqnarray}
\tilde \u_1^{(1)}(a,\theta) 
&=&
\notag
 \frac{a\omega}{2} (  \alpha  \sin\theta\cos\theta + \beta e^{i\phi} \sin \theta )+
\frac{a\omega}{3} \beta e^{i\phi}( \cos\theta \e_r - \sin\theta \e_\theta)\\
&=&
\frac{ a\omega}{6}[\alpha \tilde \u_\alpha^{(1)} (a,\theta)+ 
\beta e^{i\phi}  \tilde \u_\beta^{(1)} (a,\theta)]
,\end{eqnarray}
where we have denoted
\begin{equation}
\tilde \u_\alpha^{(1)} (a,\theta)= 3 \sin\theta\cos\theta\, \e_\theta
,\quad
\tilde \u_\beta^{(1)}(a,\theta)=2\cos\theta\, \e_r + \sin\theta\, \e_\theta
.\end{equation}
The solution to the Stokes flow problem at first order with these boundary conditions is given by  Blake \cite{Blake1971a}  and we obtain
\begin{subeqnarray}
 \tilde\u_\alpha(r,\theta) &= & \frac{3}{2}(3\cos^2\theta-1)\left(\frac{a^4}{r^4}-\frac{a^2}{r^2}\right)\, \e_r +
 3 \frac{a^4}{r^4}\sin\theta\cos\theta \, \e_\theta, \\
 \tilde\u_\beta(r,\theta)& = & 2\frac{a^3}{r^3}\cos\theta\,  \e_r + 
 \frac{a^3}{r^3}\sin\theta\, \e_\theta.
\end{subeqnarray}

\subsection{Non-Newtonian squirming}
With this solution we can then compute the non-Newtonian term in Eq.~\eqref{U2_sphere}. Rewriting Eq.~\eqref{U2_sphere} as
\begin{eqnarray}
\langle \U_2 \rangle
 & = & \langle \U_2 \rangle_N + \langle \U_2 \rangle_{NN}
\end{eqnarray}
Above we computed
\begin{equation}
 \langle \U_2 \rangle_N = \frac{4 \alpha \beta }{15} \ a \omega \sin\phi\,\e_z
\end{equation}
and recall that  we have from the integral theorem
\begin{eqnarray}
\langle \U_2 \rangle_{NN}=-\frac{1}{6\pi\eta a}\int\!\!\!\int\!\!\!\int_{V_0}\langle \BS[\u_1]\rangle :  \hat\E(\hat \P) \, \d V.
\end{eqnarray}
An explicit calculation for the integrand  exploiting Eq.~\eqref{103} leads to the final result
\begin{equation}\label{U2NN_s}
\langle \U_2 \rangle_{NN}=a\omega \frac{\alpha\beta}{15} \left[\frac{(\cos \phi  + 4 \De_1 \sin \phi )(\De_2-\De_1)}
{\De_1^2 + 1}\right] \,\e_z,
\end{equation}
{where we have defined the two Deborah numbers for the flow, 
$\De_1=\lambda_1 \omega$ and $\De_2=\lambda_2 \omega$}. 
The ratio  between the of magnitudes of non-Newtonian and Newtonian velocities is given by
\begin{equation}\label{ratioU}
\frac{\langle U_2 \rangle_{NN}}{ \langle U_2 \rangle_N}=
 \frac{(\cos \phi  + 4 \De_1 \sin \phi )(\De_2-\De_1)}
{4 \sin\phi(1+\De_1^2)}\cdot
\end{equation}

The results of Eq.~\eqref{U2NN_s} and Eq.~\eqref{ratioU} can be used to obtain a number of interesting conclusions. First, we can pick the value of the phase, $\phi$, which will lead to reciprocal motion (physically, a standing wave of actuation along the swimmer surface), 
$\sin \phi=0$. This leads to   $\langle U_2 \rangle_N=0$ while $\langle U_2 \rangle_{NN}\neq 0$, indicating, as announced in \S\ref{sec:scallop},  that an Oldroyd-B fluid can be used to induce reciprocal swimming. 

For a phase $\phi=\pi/2$ where the two surface modes are completely out of phase, we then obtain a ratio
\begin{equation}\label{phipi2}
\frac{\langle U_2 \rangle_{NN}}{ \langle U_2 \rangle_N}=
 \frac{  \De_1  (\De_2-\De_1)}
{ 1+\De_1^2 }\cdot
\end{equation}
This  is identical to the small-amplitude  result for Taylor's swimming sheet in a viscoelastic fluid \cite{lauga07} whose kinematics are that of a traveling wave. Indeed a traveling wave of the form $\cos (kx-\omega t)$ can be interpreted as the linear  superposition of two standing waves out of phase with each other.  Since {we always have $\lambda_2 < \lambda_1$}, this means that $\De_2<\De_1$, and therefore      the ratio ${\langle U_2 \rangle_{NN}}/{\langle U_2 \rangle_N}$ in Eq.~\eqref{phipi2} is negative, indicating that in this case viscoelastic stresses slow down the swimmer. By comparing the total swimming velocity to the Newtonian one we obtain in this case
\begin{equation}\label{total}
\frac{\langle U_2 \rangle_{N}+\langle U_2 \rangle_{NN}}{ \langle U_2 \rangle_N}=
 \frac{1+  \De_1  \De_2}
{ 1+\De_1^2 },
\end{equation}
and thus non-Newtonian swimming occurs always in the same direction as its Newtonian counterpart, but with a decreased magnitude.

Thirdly, we see by taking the limit of Eq.~\eqref{ratioU} for large values of $\De$ that  
\begin{equation}\label{limit}
\lim_{\De\to\infty} \frac{\langle U_2 \rangle_{NN}}{ \langle U_2 \rangle_N}= \frac{  \De_1  (\De_2-\De_1)}
{ 1+\De_1^2 },
\end{equation}
which is the same result as Eq.~\eqref{phipi2} (and  Eq.~\eqref{total} remains valid in this limit). Independently of the phase, at high Deborah number the swimming speed  always ends up being decreased by viscoelasticity.

Finally, we can use Eq.~\eqref{ratioU} to obtain a class of Newtonian swimmers whose propulsion speeds are increased by the presence of viscoelasticity. To obtain increase swimming we need $\langle U_2 \rangle_{NN}$ and $\langle U_2 \rangle_{N}$ to be of the same sign, and thus from Eq.~\eqref{ratioU} we see that this is equivalent to the mathematical requirement
\begin{equation}\label{cot}
 \cot \phi  
 < -4 \De_1.
\end{equation}
For a fixed  value of $\De_1$, we can find  values of the phase between 0 and $2\pi$ which satisfy Eq.~\eqref{cot}, leading thus to enhanced swimming at that Deborah number.  Since $\langle U_2 \rangle_{NN}$ is zero for zero Deborah number and since we have 
the asymptotic result of Eq.~\eqref{limit} at large values, we would obtain a maximum of the swimming speed at an intermediate value of  Deborah numbers in this case. In fact, a small-$\De$ expansion of Eq.~\eqref{ratioU} shows that 
\begin{equation}
\frac{\langle U_2 \rangle_{NN}}{ \langle U_2 \rangle_N}\sim 
 \frac{\De_2-\De_1}
{4 \tan \phi} + O(\De_1^2, \De_1\De_2),
\end{equation}
and thus we will obtain a range of Deborah numbers with enhanced viscoelastic swimming    in all cases where $\tan\phi< 0$. The critical Deborah number beyond which viscoelasticity always decreases swimming is given by Eq.~\eqref{cot}.

\section{Conclusion}

In this paper we  derived three general integral theorems to quantity the locomotion of isolated swimmers in non-Newtonian fluids {by  adapting classical work on the transport of small particles in non-Newtonian flows to the case of self-propulsion}. The first theorem was valid for squirmers undergoing purely tangential deformation in linearly viscoelastic fluids, and in that case the swimming kinematics were obtained to be identical to the Newtonian case. The second theorem was valid for large, arbitrary, swimmer deformation but assumed small viscoelastic behavior, for example a small Deborah number for a viscoelastic fluid or small Carreau number for a generalized Newtonian flow. The final theorem allowed order-one Deborah number but assumed that the deformation was time-periodic and of small-amplitude. That third derivation, significantly more lengthy but  more general than the previous two, exploited results of vector calculus and differential geometry to obtain a final integral formula valid for a  wide class of non-Newtonian and surface-deformation models. In all three cases, the final integrals require at most the 
mathematical knowledge of a series of Newtonian flow problems, and will be useful to quantity the locomotion of biological and synthetic swimmers in complex environments.

Our results were then  used to  show that, generically, the scallop theorem should not be expected to hold in the presence of non-Newtonian stresses. An explicit example of a swimmer unable to move in a Newtonian fluid but swimming in presence of elastic stresses in an Oldroyd-B fluid was derived. We further demonstrated that there was no {\it a priori} relationship between the direction and magnitude of the non-Newtonian and Newtonian components  of the  swimming kinematics. Specific examples were derived where small-amplitude Newtonian locomotion  could be either enhanced or decreased in an Olrdoyd-B fluid. Past experimental and computational results are therefore not necessarily  in contradiction with each other, and changing kinematics or rheological properties can qualitatively impact the non-Newtonian influence on swimming. Future computational work will be  necessary to fully untangle the relative effects of elastic vs.~shear-dependent stresses.

{Furthermore, and in the same way that our work was inspired by classical derivations on the  motion of solid particles,  the results in our paper  could be adapted to address the migration of particles in oscillatory shear flows where recent experiments \cite{lormand04} and numerical simulations \cite{davino10}  under confinement have shown interesting dynamics, including an instantaneous inversion of the direction of the wall-induced force at high frequencies as well as the presence of dead zones with very little viscoelastic migration.}

\section*{Acknowledgements}
We thank Gwynn Elfring for critical feedback on the work in \S\ref{sec:small}. This work was funded in part by the European Union via a Marie Curie CIG grant.

\bibliographystyle{unsrt}
\bibliography{theorems}
\end{document}